\documentclass[reprint,aps]{revtex4-2}

\usepackage{xcolor}
\usepackage{graphicx}
\usepackage{amsmath}
\usepackage{amssymb}
\graphicspath{{plot/}}

\usepackage{hyperref}
\hypersetup{
    colorlinks = true,
    linkcolor = blue,
    anchorcolor = blue,
    citecolor = magenta,
    filecolor = blue,
    urlcolor = blue
}
\newcommand{\SK}[1]{\textcolor{black}{{#1}}}

\begin{document}
 
\title{Scaling Description of the Relaxation Dynamics and Dynamical Heterogeneity of an Active Glass-forming Liquid}
\author{Subhodeep Dey}
\author{Smarajit Karmakar}
\email{smarajit@tifrh.res.in}
\affiliation{
Tata Institute of Fundamental Research, 
36/P, Gopanpally Village, Serilingampally Mandal, Ranga Reddy District, 
Hyderabad, 500046, Telangana, India.}

\begin{abstract}
\SK{Active glasses refer to a class of driven non-equilibrium systems that share remarkably similar dynamical behavior as conventional glass-formers in equilibrium. Glass-like dynamical characteristics have been observed in various biological systems from micro to macro length scales. As activity induces additional fluctuations in the system, studying how they couple with density fluctuations is an interesting question to address. Via extensive molecular dynamics simulations, We show that activity enhances density fluctuations more strongly than its passive counterpart. Increasing activity beyond a limit results in the sub-Arrhenieus-type relaxation behavior in active glasses. We also propose a unified scaling theory that can rationalize the relaxation spectrum over a broad parameter range using the concept of an effective temperature. In particular, we show that our scaling theory can capture the dynamical crossover from super to sub-Arrhenius relaxation behavior by changing activity from small to large values. Furthermore, We present non-trivial system size dependencies of the relaxation time at large activity limits that have not been found in any passive systems or even in active systems at small activities.} 
\end{abstract}

\maketitle
\section{Introduction}
\SK{Active matters are ubiquitous in the realm of living systems ranging from macro to micro scales. Examples include dynamics of constituents of cell cytoplasms, collective dynamics of cells in cancer progression \cite{Kim2020} and wound healing \cite{Poujade2007, Vishwakarma2020} along with their synthetic counterparts like a dense collection of self-propelled Janus colloidal particles \cite{Jiang2010}, self-propelled rods in vibrated disks \cite{Narayan2007}, etc. The discovery of a plethora of new physical phenomena in these active matters certainly calls for a better understanding from the point of view of underlying physical mechanisms both at micro and macro scales. In this regard, minimalistic models that can mimic some of the essential dynamic characteristics found in these non-equilibrium systems will often be essential for discovering the physical principles that govern the dynamics and thermodynamics of these systems. These systems show fascinating and complex dynamical properties, as they are inherently out of equilibrium in nature, and they do not follow detailed balance \cite{Vicsek1995, Toner1998, Ramaswamy2010, Palacci2013, Marchetti2013}. The study of these systems is intriguing and intricate, as a proper statistical description of the collective dynamical behavior of these systems is still lacking and is one of the most active fields of research in the condensed matter community.} 

\SK{Active matter refers to a system where the components can move internally due to their internal energy, in addition to being influenced by thermal fluctuations in the environment. Active systems exhibit a wide range of interesting dynamical phenomena, including spontaneous symmetry breaking of the rotational order in two dimensions, leading to the formation of ordered phases of clusters or flocks \cite{Vicsek1995, Toner1998}. These clusters have coherent collective motion at low noise strength and high particle density. Many biological systems exhibit collective dynamical behavior, in which forces generated by ATP consumption drive the dynamics instead of thermal fluctuations. A simple model of these systems that can capture some of the salient dynamical behaviors is a collection of self-propelled particles (SPPs) \cite{Marchetti2013}.}

\SK{Active glasses, on the other hand, represent a category of materials comprising self-propelled particles (SPPs) that display glass-like behavior characterized by slow and heterogeneous dynamics \cite{Mandal2016, Paul2023, Dey2022, Berthier2019, Janssen2019, Sadhukhan2024}. There are a lot of studies in the recent literature that show that collective dynamics of cells and tissues during cancer progression, cell proliferation, and wound healing \cite{Zhou2009, Angelini2011, Parry2014, Park2015, Garcia2015, Malinverno2017, Nishizawa2017, Kim2020, Vishwakarma2020, Cerbino2021} have dynamical features that are very similar to glass-like dynamical behaviors. The steady-state dynamical properties of many model active systems have been studied using equilibrium statistical mechanics within the ambit of linear response behavior, at least in the small activity limit via an effective temperature-like description. This effective description of the non-equilibrium steady state behavior of the system is often found to be a useful description of the system's dynamics \cite{Fodor2016}, suggesting that the effect of activity for some observables can be very well understood by assuming activity to be yet another source of noise which has temperature-like behavior at a coarse-grained time and length scales. Analytical results on the dynamics of an active particle in a harmonic potential,  known in the literature as the active Ornstein-Ullehnbeck process (AOUP), suggest that effective temperature-like dynamics accurately describe the dynamical process of the active particle \cite{Fodor2016}. However, higher-order dynamical correlation functions like four-point susceptibility ($\chi_4(t)$, defined later) are found to be not trivially understandable within the same effective temperature description \cite{Paul2023}. Thus, a clear understanding of active systems in their dynamical steady states still needs to be improved as the intricate effects of active driving on the dynamics continue to puzzle the scientific community.}

\SK{This manuscript is organized as follows. First, we discuss the details of the model glass-forming liquid and the simulation protocols employed in this work. Then, we present a scaling theory that explains the relaxation behavior of the system over the entire parameter ranges that are studied in this work, especially crossover from glassy regimes with super-Arrhenius relaxations to non-glassy sub-Arrhenius behavior with increasing activity. We then discuss novel finite-size effects observed in the relaxation time of the system at high activity limits and try to understand that using detailed finite-size scaling analysis. Next, we discuss the dynamical heterogeneity in the system both at small and large activity limits and compute the long-range dynamical correlations via displacement-displacement correlation function, performing detailed finite size scaling of the four-point susceptibility. Finally, we conclude with the implications of these results on the existing understanding and discuss the importance of scaling theories in providing insights for these non-equilibrium systems until a detailed microscopic theory is developed.}

\section{Models and methods}
\SK{\noindent{\bf Model and Simulation Details: }In this work, we studied the dynamics of a model glass-forming liquid via extensive computer simulations. We have used the well-known Kob-Anderson model \cite{Kob1995} (referred to here as 3dKA), a Binary glass-forming model system with a number ratio of ($A:B$) $80:20$, Where the $A$ or $B$ type particles are the large or small particles, respectively. The particles interact in the system according to the well-known Lennard-Jones (LJ) potential, and the interaction has been smoothed such that the 2nd derivative of the potential is zero at the cutoff distance $r_c$. The potential is given as 
\begin{equation}
    \phi(r) = \begin{cases}4\epsilon_{\alpha\beta}\left[\left(\frac{\sigma_{\alpha\beta}}{r}\right)^{12} - \left(\frac{\sigma_{\alpha\beta}}{r}\right)^{6} + c_0 + c_2r^2 \right] &, r<r_c \\
    0 &, r \geq r_c.
    \end{cases}
\end{equation}}
\SK{Here, {$\alpha, \beta$} is the type of particles, large (A) or small (B). The interaction energy between pairs of particles are $\epsilon_{AA}=1.0$, $\epsilon_{AB}=1.5$, $\epsilon_{BB}=0.5$ and the interaction diameter of the particles are $\sigma_{AA}=1.0$, $\sigma_{AB}=0.8$, $\sigma_{BB}=0.88$. The interaction cut-off distance $r_c=2.5\sigma_{AB}$. The reduced unit of energy, distance, and time are given by $\epsilon_{AA}$, $\sigma_{AA}$ and $\sqrt{\frac{\sigma_{AA}^2}{\epsilon_{AA}}}$. The integration step size is $\delta t = 0.005$, and the number density of the system is fixed at $\rho = 1.2$ for all the cases. We have performed a large-scale simulation of system size $N \in [500, 25000]$ in this work. We ran $32$ statistically independent ensembles for systems ranging below $N = 25000$ and $8$ ensembles for $N = 25000$.} 

\SK{\noindent{\bf Modelling Activity: }The activity in the system is introduced in the form of run and tumble particle (RTP) dynamics \cite{Mandal2016, Paul2023, Dey2022}, where the dynamics of the active particles can be tuned using three parameters such as concentration of active particles ($c$), force per active particle ($f_0$), and persistent time of the active particles ($\tau_p$). We solved Newton's equation of motion in the presence of active forces, and the modified equations of motion are
\begin{eqnarray}
\dot{\vec{r}}_i&=&\frac{\vec{p}_i}{m}\nonumber\\
\dot{\vec{p}}_i &=& -\frac{\partial \phi}{\partial \vec{r}_i} + \Theta_i\vec{F}_i^A,
\end{eqnarray}
with $\vec{r}_i$ and $\vec{p}_i$ being the position and momentum vector of $i^{th}$ particle, $\Theta_i$ is the active-tag which take values $1$ or $0$ depending on whether the particles is active or passive, $\phi$ is the inter-particle potential, and $\vec{F}_i^A$ is the active force. The active force on the $i^{th}$ particle in 3D can be written as  
\begin{equation}
    \vec{F}_i^A = f_0(k_x^i \hat{x} + k_y^i \hat{y} + k_z^i \hat{z}),
\end{equation}
where ($k_x,k_y,k_z$) is chosen from $\pm 1$, such that $\sum_{\alpha,i} k_{\alpha}^i=0$, i.e., the net active momentum along any direction is zero. Unless otherwise mentioned explicitly, we kept $\tau_p = 1$, active force magnitude $f_0$ is selected from $0.0 - 5.0$ for 3dKA by fixing concentration $c=0.1$. Again, to check the influence of the large activity, we have also tuned the concentration of active particles from $0.0 - 0.6$ by fixing $f_0=1.0, 2.0$. In the cases where we varied the persistence time $\tau_p$, we kept $c = 0.1$ and $f_0 = 2.0$ fixed. For getting the system size effect, we have fixed the temperature of the system of size $N=10^3$ for a given activity such that the relaxation of the system remains around $10^3$. Unlike the conventional ABP (Active Brownian particle) model, this RTP model preserves the contribution of the system's inertial effect, which holds additional information about the system's intrinsic properties. Recent work also suggests that the inertial term is essential to understanding active matter systems \cite{teVrugt2023}. We used a three-chain No\'se-Hoover thermostat to perform NVT simulations \cite{Martyna1992, Martyna1996}.}

\SK{
\noindent{\bf Dynamical Correlation Functions: } We measured two point density correlation function $Q(t)$ to compute the relaxation time of the system. $Q(t)$ is defined as 
\begin{equation}
 Q(t) = \frac{1}{N} \sum^N_{i=1} \theta(|\vec{r}_i(t)-\vec{r}_i(0)|),
\end{equation}
where $\theta(x)$ is a window function and is $1$ for $x<a$ and $0$ otherwise. $\vec{r}_i(t)$ is the position of the $i^{th}$ at time $t$. The coarse-graining parameter `$a$' is chosen to remove possible decorrelation happening due to the vibration of the particles inside the cage formed by other particles in these dense systems. We choose the value of '$a$' from the plateau of a `mean-square displacement' (MSD) in the supercooled temperature regime. For all cases, $a$ is set to $0.3$. From this two-point correlation function, we define a relaxation time $\tau_{\alpha}$, as $\left<Q(t=\tau_{\alpha})\right>=1/e$, where $\left<...\right>$ denotes ensemble average and time origin average and $e$ is the base of the natural logarithm. We equilibrated the samples for more than $100\tau_{\alpha}$, then ran the simulations for another $150\tau_{\alpha}$ for our measurements.} 
 
\SK{The four-point correlation function, $\chi_4(t)$, measures the fluctuation of the two-point correlation function $Q(t)$ and it is a well-known quantifier for measuring the dynamic heterogeneity in the system\cite{Dasgupta1991, Karmakar2009}.  $\chi_4(t)$ is defined as,
\begin{equation}
 \chi_4(t) = N \left[\left<Q(t)^2\right> - \left<Q(t)\right>^2\right],
\end{equation}
where $\left<...\right>$ refers to the ensemble average as well as the time origin average. Dynamic heterogeneity broadly refers to the heterogeneous dynamical relaxation processes in various parts of the system. This happens due to different populations of slow and fast-moving particles in the system. DH reaches its maximum around the relaxation time $\tau_{\alpha}$ and the peak is defined as $\chi_4(t=\tau_{\alpha}) \simeq \chi^P_4$. 
}

\begin{figure*}[!htpb]
\centering
\includegraphics[width=1.01\textwidth]{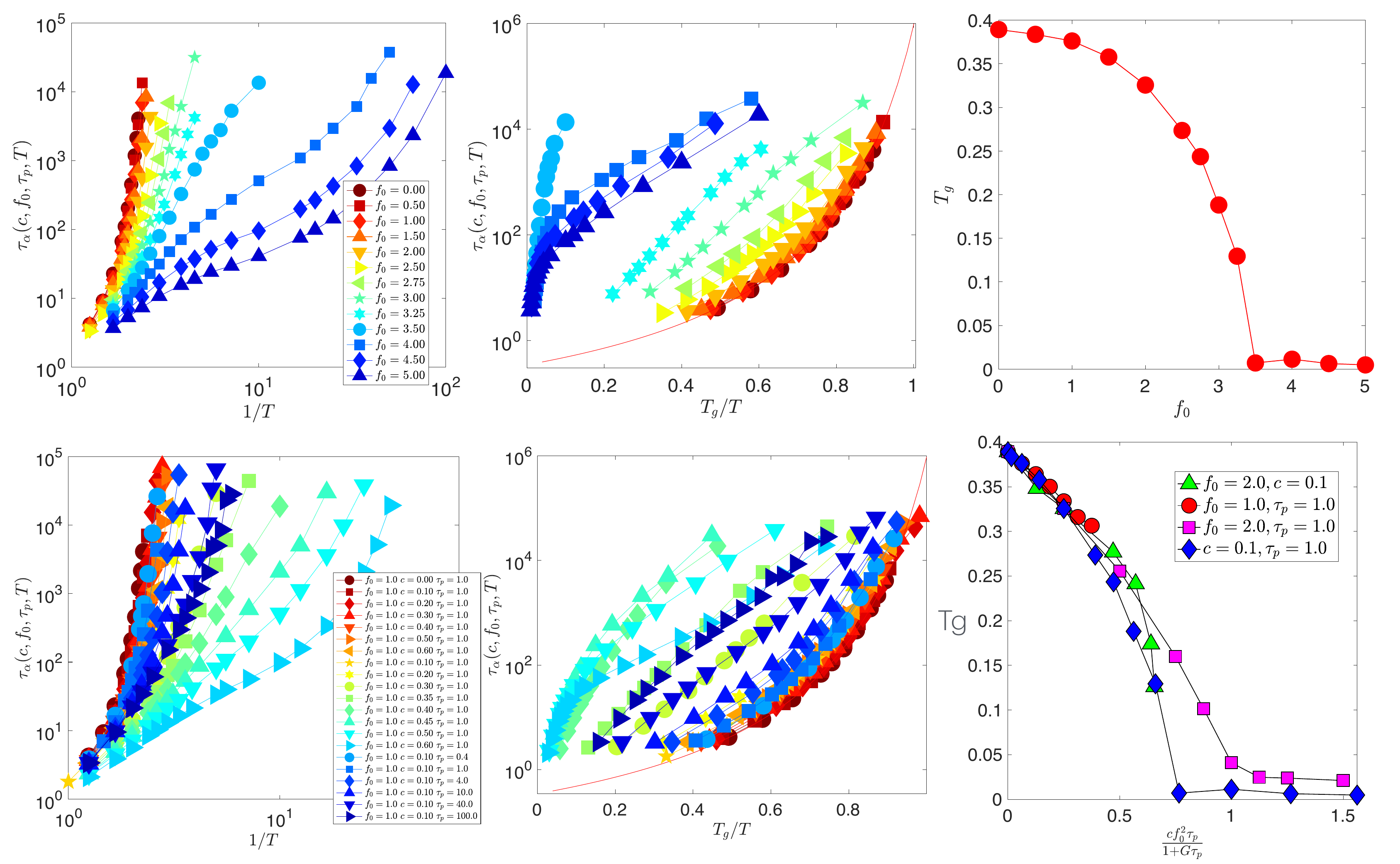}
\caption{\textbf{Angell plots}:(a) Relaxation time ($\tau_{\alpha}$) is plotted against $1/T$ for different activity $f_0$ for fixed concentration of active particles $c = 0.1$. (b) Angell plot for different activity $f_0$. Here $T_g$ is the temperature at which the system will have relaxation time $\tau_{\alpha}=10^6$ following the VFT formula. (c) The estimated $T_g$ of the active system for changing activity $f_0$. This shows that for $f_0 > 3.5$, the system does not have a glass-like phase at low temperatures. (d) Shows a similar plot of $\tau_\alpha$ as a function of $1/T$ over the studied range of parameters $f_0$, $c$, and $\tau_p$. (e) shows the Angell plot for all the data. (f) Variation of $T_g$ as a function of effective activity parameter $\Omega = cf_0^2\tau_p/(1 + G\tau_p)$ with $G = 0.6$ (see text for further details). Note that $T_g$ is very well described by an effective activity parameter up to a value, and beyond that, it shows that variation of various activity parameters can not be unified probably by a single effective parameter. }
\label{fig:angellPlot}
\end{figure*}

\section{Results}
\noindent\SK{{\bf Relaxation Dynamics: }First, we discuss the relaxation time and its dependence on $f_0$ and $c$ and $\tau_p$ both in the small and large activity limits. Fig. \ref{fig:angellPlot} (a) shows the temperature dependence of relaxation time, $\tau_{\alpha}$, for different active force $f_0$ keeping $c = 0.1$ and $\tau_p = 1.0$ constants. One can see that for smaller activities, the relaxation time shows super-Arrhenius temperature dependence, and as activity increases, one sees a strong departure from that, eventually leading to sub-Arrhenius behavior. Often, in the supercooled temperature regime, the diverging nature of the relaxation time at lower temperatures can be very well described by the Vogel-Fulcher-Tammann (VFT) relation given by Eq. \ref{eq:VFT},
\begin{equation}
        \tau_{\alpha} \simeq \tau_0 \exp\left[\frac{1}{K \cdot (T/T_{VFT}-1)}\right]\label{eq:VFT},
\end{equation}
with $K$ denoting the kinetic fragility of the system and $T_{VFT}$ being the extrapolated temperature where relaxation time will diverge. In our study, we will use a slightly modified version of this relation, which is dictated by the scaling theory discussed below. We will use the following modified VFT relation given in Eq. \ref{eq:VFTmod} to describe our relaxation time data. Note that this choice does an equally good job of describing the data at low temperatures. The modified VFT relation used here is 
\begin{equation}
        \tau_{\alpha} \simeq \tau_0 \exp\left[\frac{1}{K \cdot (T/T_{VFT}-1)}\right]^{\delta}\label{eq:VFTmod}.
\end{equation}
with $\delta \simeq 1.5$ for all the temperatures and activity parameters varied in this work. For a particular activity, the VFT relation can be used to fit the relaxation time in the supercooled regime to obtain the fragility parameter ($K$), VFT-temperature ($T_{VFT}$), and $\tau_0$. By extrapolating the VFT-relation for $\tau_{\alpha}$ using the same fitted parameter, we can get the calorimetric glass transition temperature ($T_g$) using the relation $\tau_{\alpha}(T = T_g) \sim 10^6$.} 

\SK{In Fig. \ref{fig:angellPlot} (b), we plotted the relaxation time as a function of scaled inverse temperature $T_g/T$; this representation of data is well-known in the literature as the "Angell-plot" \cite{Angell1995}. It gives a direct visualization of the fragility of the system. The one that shows the nearly straight line behavior in this figure is called the ``strong" liquid, and the one that shows a deviation below the straight line is called the ``fragile" liquid. These fragile liquids show strong super-Arrhenius temperature dependence on the relaxation time. On the other hand, the relaxation curves that show a deviation above the straight line are the ones that show sub-Arrhenius temperature dependence. A naive fit of the same VFT relation to these relaxation curves will result in negative $T_{VFT}$ as well as negative fragility index $K$. Fig. \ref{fig:angellPlot} (c) shows that the fragility ($K$) decreases with increasing activity, eventually going to a sub-zero value, indicating that at higher activity regimes, the relaxation behavior is sub-Arrhenius in nature. From Fig. \ref{fig:angellPlot} (b), one can infer that the system behaves as a fragile glass-former for $f_0<3.25$ with relaxation time following super-Arhenius behavior, but for $f_0>3.25$ the behavior completely changes to sub-Arrhenius. This suggests that by using activity alone, one can make a system transition from super-Arrhenius to sub-Arrhenius behavior. This similar transition can be observed in a passive system by changing the density of the system \cite{Berthier2009EPL, Berthier2009PRE, Adhikari2023}.}

\SK{In Fig. \ref{fig:angellPlot} (f), we have plotted $T_{G}$ for various choices of activity parameters ($f_0$, $c$ and $\tau_p$), and one finds that effective activity parameter, $\Omega = \frac{cf_0^2\tau_p}{1 + G\tau_p}$ with $G\simeq 0.6$ (a fitting parameter, see discussion later) is valid up to $\Omega \simeq 0.65$. Above this activity, the effect of activity due to changes in $f_0$ or $c$ or $\tau_p$ bifurcate. This indicates that the effects of changing activity in the system by changing the forcing or the concentration or the persistence time of active particles start to become different in the high activity limit. To better understand the relaxation behavior of the active glass-forming liquids in the entire parameter space and especially to have a physical understanding of the super to sub-Arrhenius dynamic crossover with increasing activity, we proposed a scaling theory that rationalizes all the results on relaxation time ($\tau_\alpha$) as measured using two-point density correlation function, $Q(t)$ (see Methods section).}

\SK{We start with the following observation that at zero activity, one can fit the relaxation time as a function of temperature via a VFT-like form as shown in Eq.\ref{eq:VFTmod}. For various degrees of activity, we use the same equation but now use an effective temperature description. In Ref.\cite{Paul2023}, it was shown that the relaxation time and its dependence on activity and temperature could be very well described by an effective temperature description as long as the degree of activity is still low and the system shows glass-like dynamical behaviour. It was not immediately clear what happens when one increases the activity to such a level that the relaxation profile is no longer super-Arrhenius but crosses over to sub-Arrhenius via an Arrhenius relaxation profile. We generalize the dependence of effective temperature on $c,f_0$ and $\tau_p$ to have other nonlinear dependence and chose the following form for our analysis. 
\begin{equation}
T_{eff} \simeq T \left[ 1 + A \left(\frac{1}{T}.\frac{cf_0^2\tau_p}{1 + G\tau_p}\right)^{\beta} +  B \left(\frac{1}{T}.\frac{cf_0^2\tau_p}{1 + G \tau_p}\right)  + \cdots\right], \label{Teff}
\end{equation}
or 
\begin{equation}
T_{eff} \simeq T \left[ 1 + A \left(\frac{\Omega}{T}\right)^{\beta} +  B \left(\frac{\Omega}{T}\right)  + \cdots\right], \label{Teff}
\end{equation}
with $\beta < 1$ and $A$, $B$ and $G$ are adjustable parameters. This choice of effective temperature is ad-hoc but comes from the fact that one can have non-linear behavior at high activities, and the physics may not be adequately described by a single parameter. Thus, this is simply an ansatz which may not be unique. 
}

 \begin{figure*}[!htpb]
\centering
\includegraphics[width=1.00\textwidth]{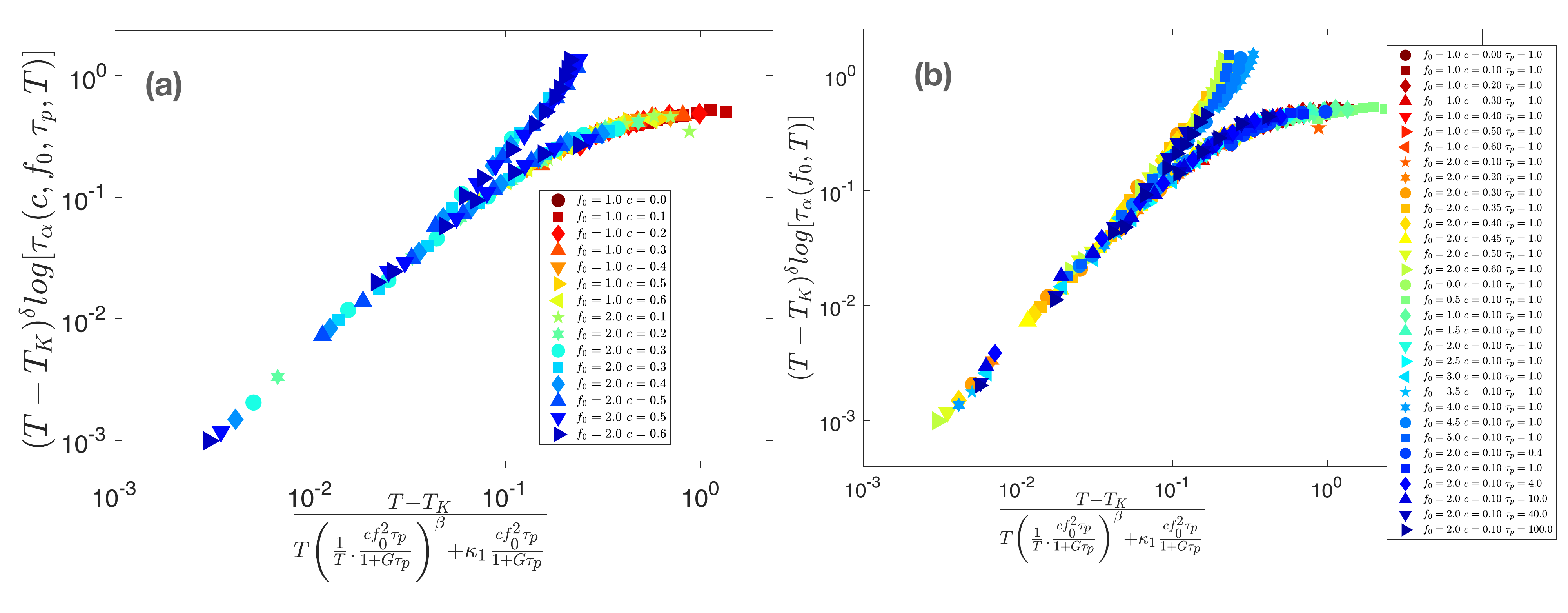}
\caption{{\bf Scaling Analysis of $\tau_\alpha$:} (a) shows the scaling analysis when the concentration of active particles varied for $f_0 = 1$ and $f_0 =2$ as discussed in the main text. The scaling collapse shows two branches. The one that shows saturation is for the situations when one gets super-Arrhenius behavior in the relaxation time. In contrast, the power-law branch is for the situations when the system shows sub-Arrhenius behavior. The scaling collapse observed is good. (b) shows similar data collapse for all the relaxation time data over the entire range of parameters studied in this work. The scaling collapse is again observed to be very good, suggesting the validity of the scaling ansatz (see text for detailed discussions).}
\label{fig:scalingAnalysis}
\end{figure*}

\SK{We start from Eq.\ref{eq:VFTmod} and use effective temperature for various activities to rewrite it as  
\begin{equation}
\tau_{\alpha}(c,f_0,\tau_p) \simeq \exp\left[\frac{1/K}{ T_{eff}(c,f_0,\tau_p)/T_{VFT}-1 }\right]^{\delta}. 
\end{equation}
Now, after substituting Eq.\ref{Teff} in this equation, we can rewrite this as
\begin{eqnarray}
|T - T_{VFT}|^{\delta}&&\log \left[ \tau_{\alpha}(c,f_0,\tau_p) \right] \simeq \\
&&\left[1 + \frac{A T \left(\frac{1}{T}\frac{cf_0^2\tau_p}{1 + G\tau_p}\right)^{\beta} +  B \frac{cf_0^2\tau_p}{1 + G\tau_p} + \cdots}{|T - T_{VFT}|}\right] \nonumber .
\end{eqnarray}
We can rewrite this equation in a scaling function form as
\begin{widetext}
\begin{eqnarray}
&&\tau_{\alpha}(c,f_0,\tau_p) \sim \exp \left [ \left(\frac{A}{|T - T_{VFT}|}\right)^{\delta} \mathcal{F}_\pm\left(\frac{|T - T_{VFT}|}{ T \left(\frac{1}{T}.\frac{cf_0^2\tau_p}{1 + G\tau_p}\right)^{\beta} + \kappa \frac{cf_0^2\tau_p}{1 + G\tau_p}  }\right) \right] \nonumber\\
&&\log{\left[\tau_{\alpha}(c,f_0,\tau_p)\right]} \sim  \left(\frac{A}{|T - T_{VFT}|}\right)^{\delta} \mathcal{F}_\pm\left(\frac{|T - T_{VFT}|}{ T \left(\frac{\Omega}{T}\right)^{\beta} + \kappa \Omega }\right).
\label{scalingEqTau}
\end{eqnarray}
\end{widetext}
by keeping only two terms and $\delta$, $\kappa$, $\beta$ and $G$ are being parameters of the scaling function. Here, the scaling function $\mathcal{F}_\pm(x)$ has the following asymptotic forms. For $x \to \infty$, $\mathcal{F}_+(x) \to \mbox{const}$ and for $x \to 0$, $\mathcal{F}_+(x) = \mathcal{F}_-(x) \sim x^\delta$. To recover Arrhenius temperature dependence at some intermediate activity, to the leading order in $cf_0^2\tau_p/(1 + G\tau_p)$, the exponent should follow the relation, 
\begin{equation}
\beta = 1 - 1/\delta.
\label{exponentRel}
\end{equation}
 All these parameters are to be determined by the data collapse to check the validity of this scaling assumption. According to Eq.\ref{scalingEqTau}, if we now plot $|T - T_{VFT}|^{\delta} \log\left(\tau_{\alpha}(c,f_0,\tau_p)\right)$ for all temperatures and activities as a function of $\left(\frac{\Omega}{T}\right)^{\beta} + \kappa \Omega $ and tune the two variables $\delta$ and $\kappa$, then one should be able to collapse all the data on master curves if the scaling ansatz is correct. In Fig.\ref{fig:scalingAnalysis}, we have shown such plots for various choices of activities. Panel (a) of this figure shows the data when we vary the active force $f_0$ keeping $c = 0.1$ and $\tau_p =1$ constants. Panel (b) shows similar results but for variation of $c$, $f_0$ and $\tau_p$ simultaneously. In the parameter regimes where the relaxation time shows super-Arrhenius-like behavior, the data will fall on the flat branch of the master curve as the scaling prefactor in the y-axis accounts for the super-Arrhenius divergence. Similarly, for the regime where the relaxation shows sub-Arrhenius behavior, the data will collapse on the power-law part of the master curve.} 

\begin{figure*}[!htpb]
\centering
\includegraphics[width=0.8\textwidth]{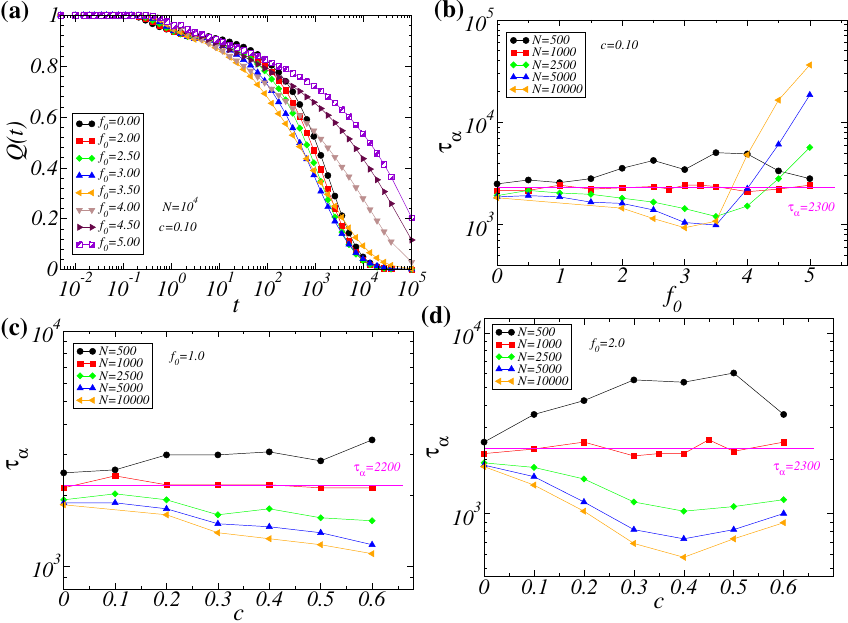}
\caption{\textbf{Relaxation behavior and Finite size effects}: (a) The two-point overlap correlation function $Q(t)$ is plotted for various activities for $N=10^4$ system size. It shows that $\tau_{\alpha}$ increases in the high activity limit when the system is in the sub-Arrhenius regime. Note that temperature for this case is chosen such that the relaxation time for $N = 10^3$ system size is nearly constant at around $2300$ (see panel b). (b) Shows variation of $\tau_\alpha$ with increasing $f_0$ for various system sizes. For smaller $f_0$ the relaxation time decreases with increasing system size, but beyond a particular activity, $f_0 \simeq 3.5$, one sees a stark difference in the system size dependence.  (c) \& (d) Variation of $\tau_\alpha$ when one varies the active particle concentration $c$ by keeping  $f_0=1.0$, and $f_0 = 2.0$ constants, respectively. Again, the non-monotonic behavior is found beyond a critical activity, indicating that the non-monotonic dependence is very generic in nature. }
\label{fig:Qvst_tauAlphavsC_f0_N_3dKA_merge}
\end{figure*}
\SK{As shown in the two panels of Fig.\ref{fig:scalingAnalysis}, it is clear that the quality of data collapse is good with only four adjustable parameters for all the data. We used $T_K = 0.295$, which is obtained by fitting the data for the passive system using the VFT equation, and this number is well-documented in the literature, so we are not considering this as a free parameter. We varied $\delta$, $\kappa$, $\beta$ and $G$ to obtain the data collapse. The best collapse is obtained using $\delta = 1.5$, $\kappa \simeq 0.75$, $\beta = 0.35$ and $G = 0.6$. Note that Eq.\ref{exponentRel} puts an additional constraint on these four variables, and thus, we have three free variables in principle. The $\delta$ and $\beta$ obtained via this scaling collapse obeys the relation Eq.\ref{exponentRel} very well. Although we have treated all these four variables as free-fitting parameters, choosing any three would have given us equally good data collapse.  It is also interesting to note that temperature dependence of relaxation time is better represented by the modified VFT equation (see Eq.\ref{eq:VFTmod}) to describe the relaxation across the parameter range in a unified manner. Interestingly, the effective temperature assumed (see Eq.\ref{Teff}) in deriving the scaling ansatz seems to suggest that the first dominating term might be sub-linear in nature as $T_{eff} = T + \kappa cf_0^2\tau_p$ does not lead to very good data collapse. Understanding why such a form of effective temperature works better across a wide range of activity will certainly be interesting for developing a better understanding of the dynamics of active glasses from a quasi-equilibrium perspective.}

\begin{figure*}[!htpb]
\centering
\includegraphics[width=1.02\textwidth]{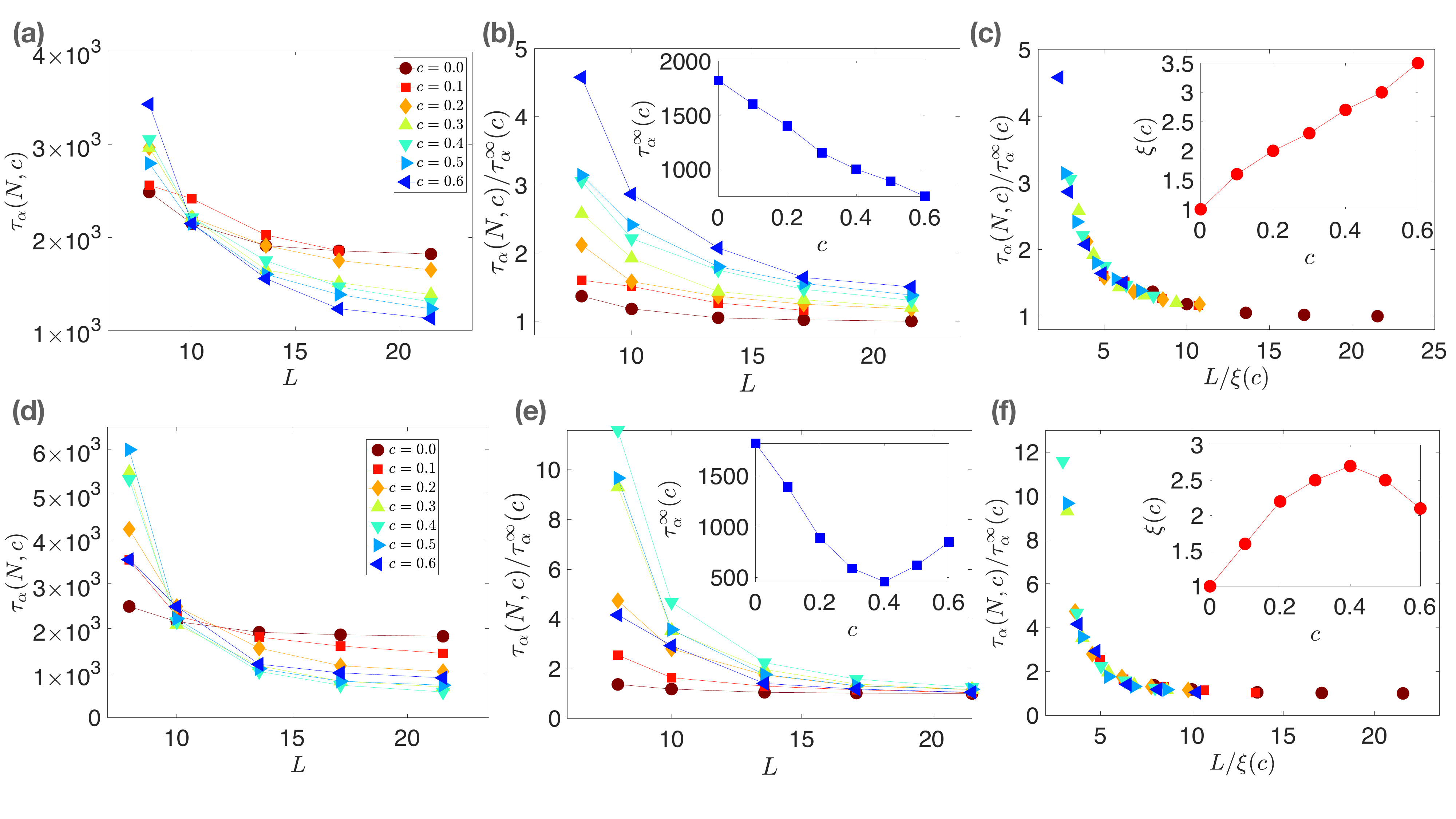}
	\caption{\textbf{Finite Size Scaling}: (a) The system size effect of the relaxation time for different concentrations given for significantly lower activity $f_0$=1.0. (b) The relaxation time for different system sizes is scaled by its asymptotic value at $L\rightarrow \infty$. In the inset, the asymptotic relaxation time ($\tau_{\alpha}^{\infty}$) is given for different concentrations, which tends to decrease monotonically with an increase in activity. (c) The system size is scaled for different activities with the static length scale, and all the data follows the universal master curve. Inset, the static length scale of the system grows systematically with increasing activity. (d) similar to plot (a), it shows the relaxation time for higher activity $f_0$=2.0. (e) shows scaled relaxation like (b) at higher activity. Inset, the asymptotic value of relation time shows the non-monotonic nature. (f) similar to plot (c), it shows the static length of the active system scaled by the static length of the passive system. The inset shows the non-monotonic nature of the static length scale. For both the non-monotonicity of ($\tau_{\alpha}^{\infty}$) and $\xi(c)$ shows the transition from super-Arrhenius to sub-Arrhenius regime.}
\label{fig:fss_c}
\end{figure*}
\SK{Now we discuss the intriguing finite-size behavior of relaxation time that contrasts with the results known in equilibrium systems. In Fig. \ref{fig:Qvst_tauAlphavsC_f0_N_3dKA_merge}(a), we show the overlap correlation function $Q(t)$ for $N = 10^4$ system size at active particle concentration, $c = 0.1$, with changing the strength of the active force, $f_0$. We see very interesting behavior which suggests that for smaller activities, the relaxation time remains very similar (we choose bath temperatures for these systems such that relaxation time $\tau_\alpha \simeq 2300$ in reduced units for $N = 10^3$ to have a possible direct comparison across changing activity) but as soon as one increases the activity, the relaxation time instead of decreasing to smaller values starts to increase which is very counter-intuitive from the perspective of an effective temperature description. A systematic fluidization at large activity is typically expected in these dense, disordered systems. These results prompted us to study the finite-size effects systematically with changing activity parameters. In Fig. \ref{fig:Qvst_tauAlphavsC_f0_N_3dKA_merge} (b), we plotted relaxation time as a function of active force, $f_0$  for various system sizes starting from $N = 500$ to $N = 10000$. One can clearly see that at a lower activity regime, the relaxation time decreases monotonically with increasing system size, eventually saturating to a constant number, whereas at a large activity limit, we can observe an increase in $\tau_{\alpha}$ with increasing system size. These results at lower temperatures are in complete agreement with the results obtained in passive systems, indicating a systematic growth of static length scale with decreasing temperature even for active systems, as shown in Ref.\cite{Paul2023}, although the growth of the length scale is much larger in active systems as compared to the passive one. On the other hand, at large activity limits, non-monotonic behavior of $\tau_\alpha$ with system size, especially at large system size, is observed. Strong growth of relaxation time is in stark contrast with passive systems, and as far as we know, it has never been reported before. A clear understanding of these results requires further exploration. Similarly, in panels (c) and (d), we show the system size dependence of relaxation time when we keep $f_0$ constants at two different values and increase the concentration of active particles. Interestingly, there is some non-monotonic behavior with changing concentration especially at large concentration the system size dependence becomes a bit weaker than at intermediate concentration. A detailed analysis of these aspects are explored in the subsequent sections.}

\begin{figure*}[!htpb]
\centering
\includegraphics[width=1.03\textwidth]{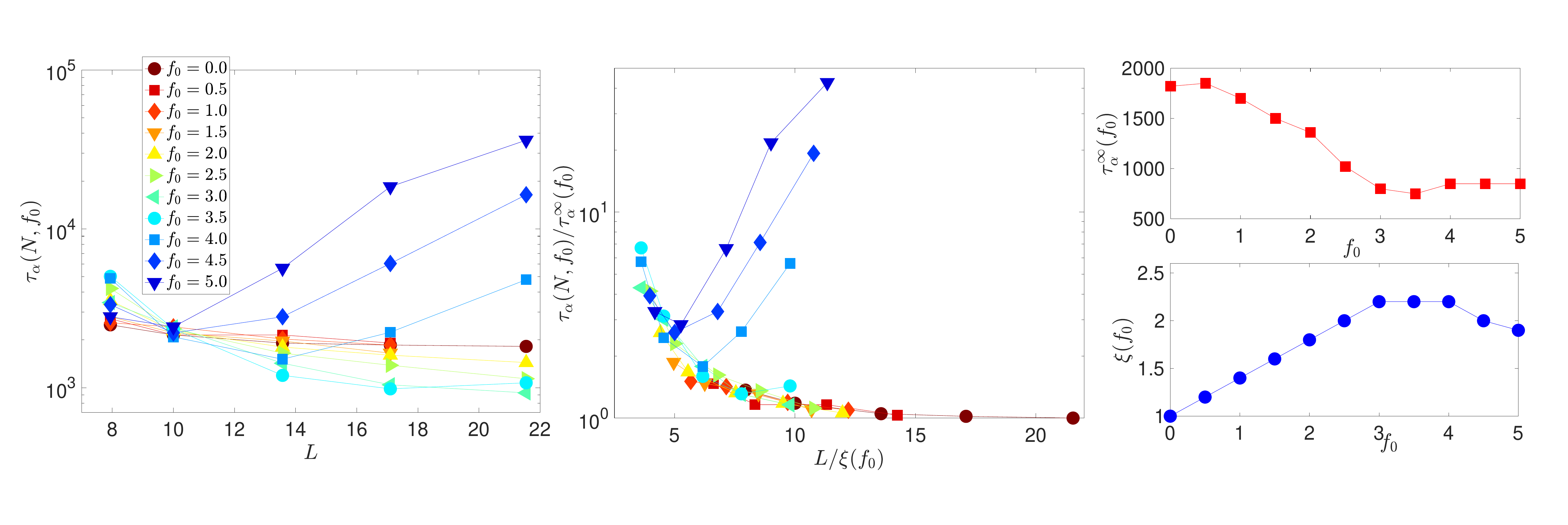}
\vskip -0.3in
\caption{\textbf{Finite Size Scaling}: (a) System size dependency of the relaxation time ($\tau_{\alpha}$) for different activity $f_0$, where the concentration of the active particles are fixed at c=0.1. Here, at a high activity limit, the relaxation time increases with the increase in system size, which is not present in any of the known passive glassy systems. (b) Now, in the large system $L \rightarrow \infty$, the $\tau_{\alpha}$ tends to reach the asymptotic limit $\tau_{\alpha}^{\infty}$. We have scaled the relaxation time using this asymptotic relaxation time for different activities $f_0$. The system size is scaled using static scale $\xi(f_0)$. It shows the scaled relaxation time with a scaled system size to follow a universal curve for a small activity limit. (c). Here, the asymptotic relaxation time decreases with increasing activity before reaching a constant value at a higher activity limit. (d) Here, the static length scale ($\xi$) for different activity $f_0$ increases for small activity limit, and at large activity limit, it reaches a constant limit.}
\label{fig:fss_f0}
\end{figure*}

\SK{To establish a link between the observed system size effect of relaxation time with changing activity, we performed a detailed finite-size scaling analysis. In Fig.\ref{fig:fss_c}, we show in the top panels the relaxation times as a function of system size, keeping $f_0 = 1.0$ and $\tau_p = 1.0$ fixed while varying the concentration of the active particles $c$. The finite size behavior in this window of activity is very similar to that observed in passive systems, albeit with stronger dependence. In panel (b), we show the scaled relaxation time as a function of $L$, and the inset shows the chosen values of $\tau_\alpha^\infty(c)$, the large system size limit values of the relaxation time. $\tau_\alpha^\infty(c)$ shows a monotonic decrease with increasing concentration of active particles. In panel (c), we show the scaling collapse where we have scaled the x-axis by an appropriate choice of the correlation length, $\xi(c)$. The correlation length shows a monotonic increase with increasing activity in agreement with the observation in Ref.\cite{Paul2023}. In the bottom panels, we show a similar set of results, but this time, we kept $f_0 = 2.0$ and varied the concentration. Interestingly, as the concentration of active particles increases, the system shows non-monotonic behavior. In panel (e), we show the scaled relaxation time with an inset showing the dependence of $\tau_\alpha^\infty$ on $c$. It shows non-monotonic dependence hinting at the increase of relaxation time with increasing activity. Panel (f) shows the final data collapse, with the inset showing the growth of the correlation length with changing concentration. The non-monotonic behavior of static length scale with the concentration of active particles is indeed very interesting, but the microscopic reasons of why the correlation length decreases with increasing activity at large activity limit are not immediately clear, although one might argue that at large activity, the system no longer shows properties of supercooled liquids with no amorphous solid phase at low enough temperatures. This corroborates well with the observation that at larger activity, the system shows super to sub-Arrhenius temperature dependence.}

\SK{Now, we discuss finite-size effects, which completely contrast with reported observations in any glass-forming liquids in equilibrium. In Fig.\ref{fig:fss_f0}(a), we show $\tau_\alpha(N,f_0)$ as a function of the linear size of the simulation box, $L$. One can clearly see that for values of $f_0$ that are low enough, the relaxation time systematically decreases with increasing system size and saturating at constant values. This is very similar to the results obtained in glass-forming supercooled liquids in equilibrium conditions, but for large active force $f_0 > 3.0$, one sees that relaxation time initially decreases with increased system size but then takes overturn and increases with increased activity very rapidly. This crossover in behavior starts to happen at smaller system sizes as the activity is increased beyond $f_0 \simeq 3.0$. Panel (b) shows a possible scaling collapse of the same data using an appropriate asymptotic timescale and a length scale. Note that only the first part, which shows a monotonic decrease with increasing system size, is being attempted to scale for larger activity data. The scaling collapse highlights the non-trivial increase in relaxation time with increasing activity for large activity very elegantly. Panel (c), (d) shows the $\tau_\alpha^{\infty}$ and $\xi$ that are used for the scaling collapse, respectively.}

\SK{This behavior, as emphasized in the previous paragraph, shows a unique scenario in which a disordered system shows an increase in relaxation time with increasing system size, similar to the critical slowing down observed in a continuous phase transition. This similarity might signal that glass-like behavior observed in high activity regimes may have dynamical behavior predicted by Mode Coupling Theory for active systems as hinted in a recent work \cite{Nandi2017}. A more detailed analysis of these results, especially at higher activity limits across different models of active glasses, will be very important to see whether the observed behavior is generic or not, and if it is generic, then an understanding of the microscopic reasons for such a crossover in the finite size behavior of relaxation time might indicate what kind of relaxation processes are at play in these different scenarios. To give an example for equilibrium systems, relaxation time in passive glass-forming liquids is known to be controlled by the configurational entropy ($S_c$), which measures the amount of accessible microstates available for relaxation and the process. The dependence is found to be exponential in nature, as predicted by Adam-Gibb's theory and the well-known activated dynamics theory of glass transition, the Random First Order Transition (RFOT) theory. These theories suggest that at lower temperatures, the relaxation process is activated in nature, whereas Mode Coupling Theory predicts that process to be similar to critical phenomena like with power-law divergence of relaxation time with decreasing temperature and a possible increase in relaxation time with increasing system size near the critical point. Thus, a crossover in finite-size effects of relaxation time might signal a crossover governing dynamics in these systems. It will be very interesting if such a connection can be established in these systems. This might pave the way for improving the current understanding of the physics of glass transition.}

\begin{figure*}[!htpb]
\centering
\includegraphics[width=0.99\textwidth]{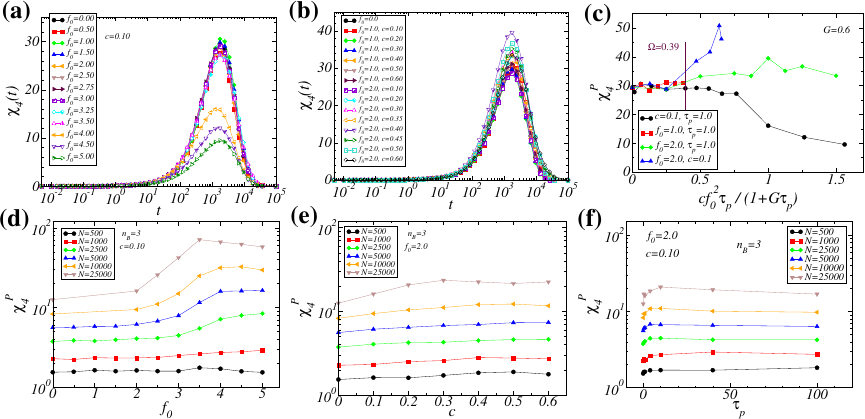}
\caption{\textbf{$\chi_4^p$ for different activity}: (a) Dynamic heterogeneity (DH) of the system of size $N=10^3$ is given for different $f_0$, where for the higher activity, it shows the decrease in DH in the canonical ensemble. Similarly, in (b), the activity is changed by tuning the active particle concentration c, where the DH peak height does not change significantly in our observed limit. (c) For $N=10^3$, in the canonical ensemble, below effective activity $\Omega$ = $[cf_0^2\tau_p/(1+G\tau_p)]$ = 0.39, the dynamic heterogeneity peak is nearly the same, and above $\Omega=0.39$ the $\chi_4^p$ tends to bifurcate due to the effect of different activities in the system, here G is chosen to be 0.6. \textbf{$\chi_4^P$ for different activity in grand-canonical ensemble}: (d) shows the increase in $\chi_4^P$ for high activity $f_0$ at different system sizes. Here, we have considered the grand canonical ensembles with sub-systems ($L_B=L/3$). Large systems tend to show a sharp increase in dynamic heterogeneity with activity, and eventually, it saturates at high activity $f_0$. Similarly, (e) $\chi_4^P$ increases with increasing activity c at fixed $f_0=2.0$ before it saturates at high activity for different system sizes. The dependence is a bit mild in this case. (f) $\chi_4^P$ increases with increasing activity $\tau_p$ at fixed $f_0=2.0,c=0.1$ before it saturates at high activity for different system sizes. Once again, the dependence is found to be a bit mild. }
\label{fig:chi4pvsC_f0_n3_N_merge}
\end{figure*}
\vskip +0.1in
\noindent{{\bf Dynamical Heterogeneity:} } \SK{Now, we focus on the dynamical fluctuations in these systems via four-point dynamical susceptibility, $\chi_4(t)$ (see definition in the Methods section).  In Fig.\ref{fig:chi4pvsC_f0_n3_N_merge}, we show $\chi_4(t)$ as a function of both increasing $f_0$, concentration $c$ and persistent time ($\tau_p$). In panel (a), we show how $\chi_4(t)$ shows mild growth up to activity $f_0 = 3.0$ and then starts to show a strong decrease with a further increase in $f_0$. Note that we have done this computation of $\chi_4$ in the canonical ensemble in which many other important fluctuations like fluctuations in density, temperature, composition, and activity are missing. Panel (b) shows similar results, but for the varying concentrations of active particles, one sees much less dependence on activity, at least for the studied parameters. In panel (c), we consolidated all the dependence of peak height of $\chi_4(t)$ (referred to here as $\chi_4^P$) and plotted them as a function of $cf_0^2\tau_p/(1 + G\tau_p)$. One sees that for smaller activity, the effective activity parameter $\Omega = cf_0^2\tau_p/(1 + G\tau_p)$ seems to describe the data reasonably well, but for large activity, the effects are very different, especially when we keep $c = 0.1$ fixed but vary $f_0$, we see a strong decrease in $\chi_4^P$, whereas when keep $f_0 = 2.0$ fixed but vary $c$, then $\chi_4^P$ starts to increase significantly which is not the case if we fix $f_0 = 1.0$ and then vary $c$. Thus, it seems that active force magnitude seems to have a much bigger role to play, especially at larger activity, than the concentration of active particles.} 

\begin{figure*}[!htpb]
\centering
\includegraphics[width=1.0\textwidth]{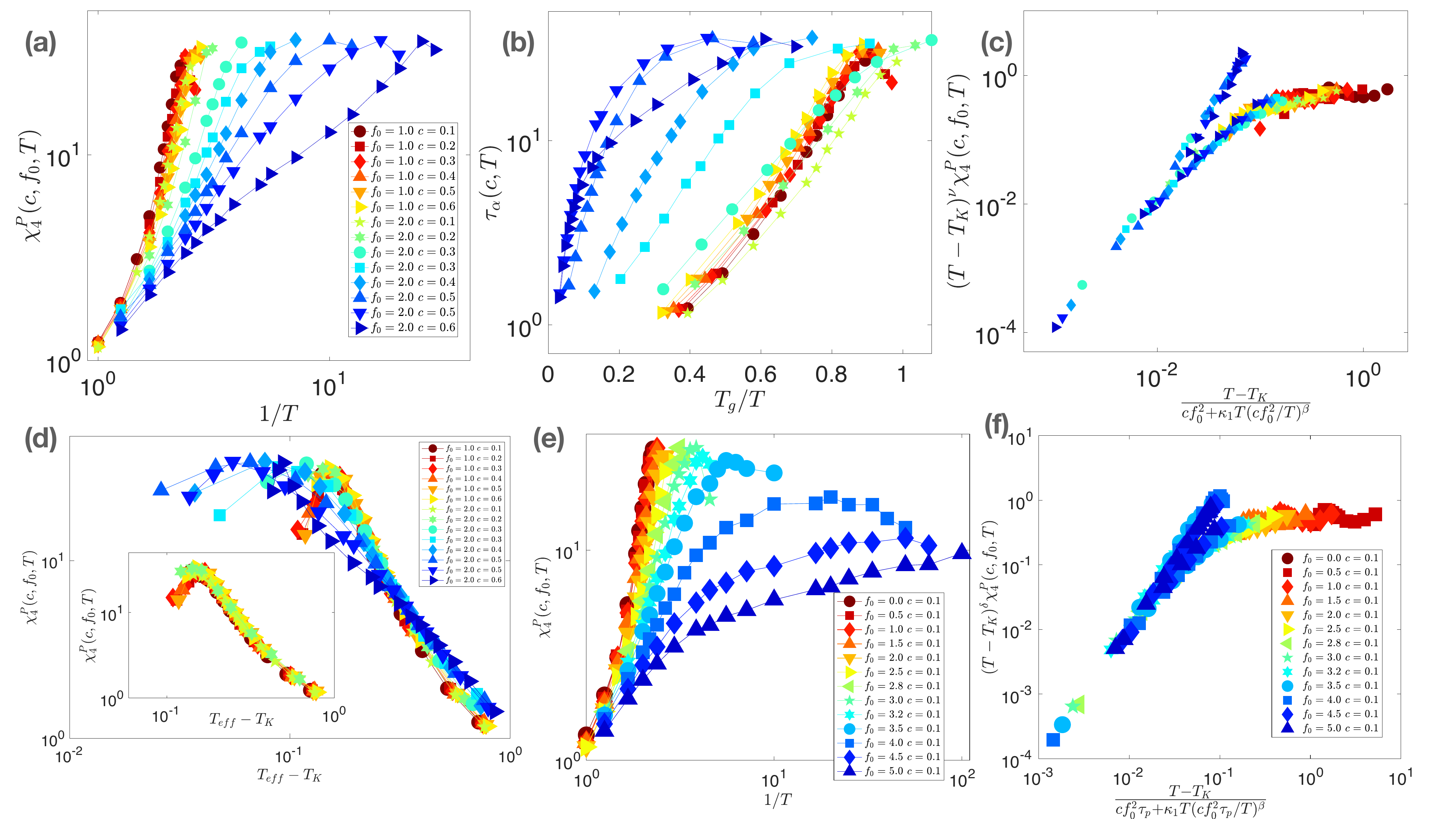}
\caption{\textbf{Scaling Analysis of Dynamic Heterogeneity}: (a)  Shows the peak of $\chi_4(t)$ defined as $\chi_4^P$ on $1/T$ for different activity.  (b)  Shows the same data, but the temperature is scaled by $T_g$. (c) Scaling analysis using scaling theory suggests that DH at a small persistence time can be described reasonably well using scaling theory. (d) $\chi_4^P$ plotted as a function of effective temperature ($T_{eff}$) in double logarithmic plot. It shows that a power-law description of the data for a small relaxation time window is not bad at all. This also suggests that mode coupling theory predictions hold reasonably well at this intermediate temperature range, similar to passive systems. The inset shows the same data but for smaller activity. The power-law description with an effective temperature is much better at this small activity window. (e) $\chi_4^P$ plotted as $1/T$ for different activity $f_0$ just to demonstrate that at high enough activity $\chi_4^P$ shows a broad maximum at intermediate temperature and shows a decrease in $\chi_4^P$ at lower temperatures. (f) Similar scaling analysis as in panel (c) but for variation of $f_0$ in the entire range of study keeping $c=0.1$ and $\tau_p = 1$ constants.}
\label{fig:chi4Scaling}
\end{figure*}

\SK{In subsequent discussions, we show the results obtained from an equivalent grand canonical ensemble in Fig. \ref{fig:chi4pvsC_f0_n3_N_merge} (d). We first divide the whole simulation box into smaller sub-boxes and then study dynamic fluctuations in these sub-boxes with linear size $L_B$. In this study, we show the results when $L_B = L/3$. This choice ensures that we have a large enough sub-box and a large enough embedding bath. In the grand-canonical ensemble, one has a very different dependence when additional fluctuations are included. $\chi_4^P$ starts to increase for larger system size, as shown in Fig. \ref{fig:chi4pvsC_f0_n3_N_merge} (d). Note that for canonical ensemble, one sees a behavior that is very different, and it suggests a decrease in fluctuations. This once again highlights how important the different contributions of fluctuations are in understanding the dynamic fluctuations in these systems. Panels (e) shows the results of keeping the active force magnitude constants at $f_0 = 2.0$ and varying the concentrations. The variation of $\chi_4^P$ with increasing concentration is not as dramatic as changing the force magnitude. Panel (f) on the other hand shows the variation of $\chi_4^P$ when $f_0 = 2.0$ and $c=0.1$ are kept constants but $\tau_p$ is varied. One sees a relatively mild dependence on $\tau_p$, eventually saturating to a constant value at large $\tau_p$ for each system sizes.  Overall, it suggests that at large enough activity limits, the dynamical fluctuations can be very different, and a dynamical description based on simple, effective temperature or effective activity parameters may not be very helpful, and a better understanding is essential.}

\begin{figure*}[!htpb]
\centering
\includegraphics[width=0.99\textwidth]{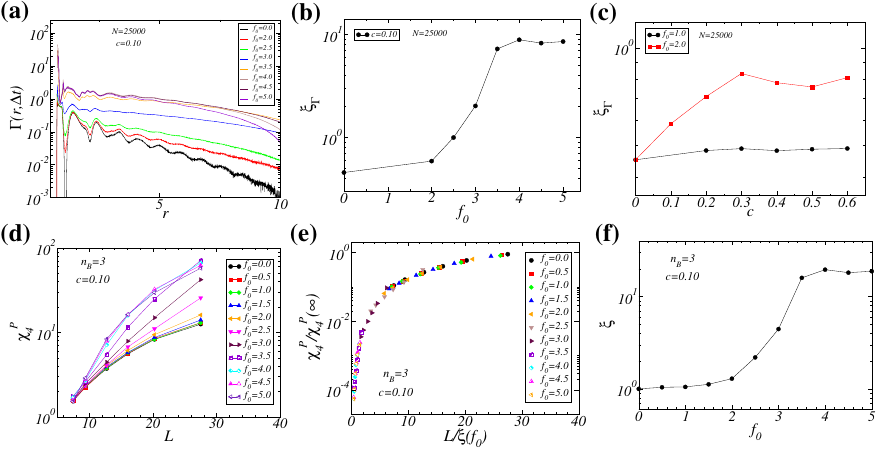}
\caption{\textbf{Growth of Correlation length}: (a) Excess displacement-displacement correlation for different activity $f_0$ at time $\Delta t$ is plotted for various $f_0$. Here, the $\Delta t$ is the time at which $\chi_4$ computed for sub-systems ($L_B=L/3$) reaches maximum value $\chi_4^P$. (b) shows the correlation length, $\xi_{\Gamma}$ of the excess displacement-displacement correlation for different activity $f_0$. It shows that the growth of the correlation length is mild at smaller activities, but then it shows a sharp increase with activity eventually saturating beyond $f_0 > 3.5$, signaling a dynamic crossover from super to sub-Arrhenius dynamics. (c) shows the correlation length for changing the concentration of active particles, $c$, for two different $f_0$. (d) Shows the block size dependence of $\chi_4^P$ for various $f_0$. with (e) showing the scaling collapse using a dynamic length scale shown in panel (f). The length scale is found to be proportional to the length scale obtained using excess displacement-displacement correlation, i.e., $\xi \propto \xi_\Gamma$. Note that the length scale obtained in any scaling analysis is unknown up to a constant prefactor.}
\label{fig:chi4pvsL_LB3_scaling_chi4p_xi_f0_3dKA_merge}
\end{figure*}

\SK{ To better understand the dynamical fluctuations, we will once again use the scaling theory approach and see whether the results can be rationalized within such an approach. We assume that for passive systems, one sees $\chi_4^P$ to obey a power-law-like temperature dependence as $\chi_4^P \sim |T - T_{VFT}|^{-\nu}$ and use a similar approximation of the effective temperature in the small $\tau_p = 1$ regime (data for large $\tau_p$ does not follow similar scaling behaviour; see discussion later), then we can write the following simplified scaling function 
\begin{widetext}
\begin{equation}
\chi_4^P(c,f_0) \sim \left(\frac{B}{|T - T_{VFT}|}\right)^{\nu} \mathcal{G}_\pm\left(\frac{|T - T_{VFT}|}{ cf_0^2 + \kappa_1 T \left(cf_0^2/T\right)^{\beta}}\right).
\label{scalingEqChi4}
\end{equation}
\end{widetext}
Thus, if the scaling function is a good description of the system, then one will be able to obtain data collapse if one plots $|T - T_{VFT}|^{\nu}\chi_4^P(c,f_0)$ as a function of $|T - T_{VFT}|/\left[ cf_0^2 + \kappa_1 T \left(cf_0^2/T\right)^{\beta}\right]$ for all the data by varying $\nu$, $\kappa_1$ and $\beta$. For the obtained collapse, we have chosen the following parameters $\nu = 2.0$, $\kappa_1 = 1.75$, and $\beta = 0.75$. In Fig.\ref{fig:chi4Scaling} (a), we show $\chi_4^P(c,f_0)$ as a function of $1/T$ with panel (b) showing the same data, but the x-axis is scaled by the estimated calorimetric temperature, $T_g$. In panel (c), we show the data collapse obtained following the scaling arguments for $f_0 = 1$ and $f_0 = 2.0$ with varying $c$. The data collapse obtained using the scaling arguments is found to be good, suggesting a possible way to rationalize the results in a unified manner. Similar data is reported in panel (f), but for $c = 0.1$ with varying $f_0$. Data collapse for these sets of data using the same adjustable parameters is found to be good. The raw data is shown in panel (e). Note that for large activities, the peak height of $\chi_4$ tends to show a non-monotonic behavior with a broad peak at an intermediate temperature regime. Panel (d) shows the data replotted as a function of $T_{eff} - T_{VFT}$ in a log-log plot to highlight that there is an intermediate temperature regime where $\chi_4^P$ shows nice power-law behavior, although at much lower temperature and high enough activities, the non-monotonic behavior becomes much clearer. Inset shows the same data but for the small activity regime ($f_0 = 1$ with $c \in [0.0, 0.6]$ and for $f_0 = 2$ with $c\in [0.1, 0.2]$. In panel (e), we have shown the variation of $\chi_4^P$ as a function of $1/T$ to better highlight the peak in $\chi_4^P$ at intermediate temperature scales for large $f_0$. In panel (f), we present the same scaling analysis but for a wide variation of $f_0$ while keeping $c=0.1$ and $\tau_p = 1.0$ constants. For this data collapse, we kept the parameters the same. The nice data collapse at intermediate temperature indeed suggests that some aspects of the physics still get explained via an effective temperature description, albeit with an approximate definition for the effective temperature.    
}

\begin{figure*}[!htpb]
\centering
\includegraphics[width=0.99\textwidth]{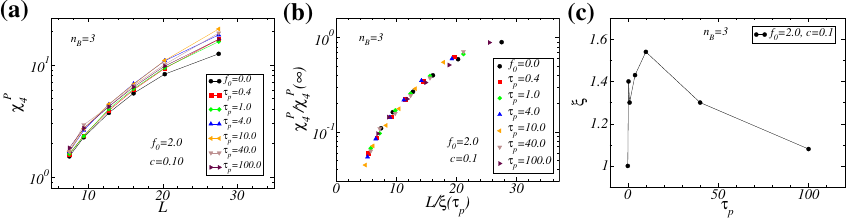}
\caption{\textbf{Effect of changing persistence time}: (a) Shows $\chi_4^P$ vs block length, $L_B$ with changing $\tau_p$ for $f_0=2.0$ and $c=0.1$. (b) Shows the scaling collapse of the same data with the correlation length shown in panel (c). Note the non-trivial dependence of the obtained correlation length with changing persistence time.}
\label{fig:largeTaup}
\end{figure*}
\SK{As we see stronger growth of dynamical heterogeneity at intermediate and high activity limits, we now focus on computing the spatial correlation function $\Gamma(r,\Delta t)$, also known as the excess displacement-displacement correlation function, to estimate the growth of underlying dynamic length scale as a function of increasing activity. $\Gamma(r,\Delta t)$ is defined as 
\begin{equation}
\Gamma(r,\Delta t) = \left[g^{uu}(r,\Delta t)/g(r)-1.0\right].
\end{equation}
}
\SK{The excess displacement-displacement correlation $\Gamma(r,\Delta t)$ at a large time would be decorrelated, and $g^{uu}$ would be equal to g(r). And $\Gamma(r,\Delta t)$ would decay to zero as a function of $r$. Now, if one assumes the correlation function decays exponentially, then the area under the curve will give us the characteristic length scale, which will be proportional to the dynamical correlation length scale $\xi_{\Gamma}$ of the system at time $\Delta t$ as shown in \cite{Poole1998} and recently demonstrated very clearly for various model glass-forming liquids \cite{Tah2020}.}

\SK{The displacement-displacement correlation function $g^{uu}(r,\Delta t)$ at time $\Delta t$ is defined as,
\begin{equation}
g^{uu}(r,\Delta t)=\frac{\left\langle  \sum\limits_{i,j=1,j\neq i}^Nu_i(0,\Delta t)u_j(0,\Delta t)\delta(r-|\textbf{r}_{ij}(0)|)\right\rangle}{4\pi r^2\Delta rN\rho \langle u(\Delta t)\rangle^2}
\label{guu}
\end{equation}
where, $u_i(t,\Delta t)=|\textbf{r}_i(t+\Delta t)-\textbf{r}_i(t)|$, and $\langle u^2(\Delta t)\rangle=\langle\frac{1}{N}\sum_{i=1}^N
\textbf{u}_i(t,\Delta t).\textbf{u}_i(t,\Delta t)\rangle$. The radial distribution function $g(r)$ is defined as,
\begin{equation}
g(r)=\frac{\left\langle\sum\limits_{i,j=1,j\neq i}^N\delta(r-|\textbf{r}_{ij}(0)|) \right \rangle} {4\pi r^2\Delta r N\rho}
\end{equation}}

\SK{
	In Fig. \ref{fig:chi4pvsL_LB3_scaling_chi4p_xi_f0_3dKA_merge}(a), we show $\Gamma(r,\Delta t)$ for $N = 25000$ particles keeping $c = 0.1$ and varying $f_0$ over the entire studied range. One can clearly notice that displacement correlation systematically becomes decorrelated over a longer spatial distance, and the corresponding correlation length $\xi_\Gamma$ is shown in panel (b). Notice that for smaller $f_0$, the growth is mild, but then it increases significantly, eventually reaching a saturation value beyond $f_0 = 3.5$ or so.} 

\SK{The growth of correlation length with $f_0$ has a strong similarity with the growth of $\chi_4^P$ for $N = 25000$ system size as a function of $f_0$, suggesting a direct relation between them in agreement with the passive systems even though these systems are in non-equilibrium.  Fig. \ref{fig:chi4pvsL_LB3_scaling_chi4p_xi_f0_3dKA_merge}(c) shows the growth of correlation length for two different $f_0$ as a function of $c$ and once again this shows a nice correspondence with the growth of $\chi_4^P$ as shown in Fig.\ref{fig:chi4pvsC_f0_n3_N_merge}. We then ask the question of whether the same length scale will control the finite-size behavior of $\chi_4^P$ or not. In Fig. \ref{fig:chi4pvsL_LB3_scaling_chi4p_xi_f0_3dKA_merge}(d), we show the system size dependence of $\chi_4^P$ computed at a fixed block length of $L_B = L/3$ and plotted the same for various $f_0$'s keeping $c = 0.1$. Again, the motivation for choosing a sub-system instead of the full system is to accommodate all the possible fluctuations that are important for the correct estimation of $\chi_4(t)$. Panel (e) shows the finite-size scaling collapse of the same data with the corresponding correlation length shown in panel (f). The length scale for different activities, $\xi_{\Gamma}(f_0)$, is reported by scaling the length scale by its value for the passive system, $\xi_{\Gamma}(0)$. The remarkable similarity suggests that these two length scales are the same. Note that in finite-size scaling analysis, the correlation length can be obtained up to an arbitrary scaling factor, but the functional dependence on the controlled parameter is free from any arbitrariness, unlike in a fitting procedure. Some of these results agree with the results reported in Ref.\cite{Paul2023}, but the saturation of $\chi_4^P$ with further increase in activity and non-monotonic system size dependence of relaxation times are yet to be understood.
}

\SK{Next, we quickly discuss the effect of increasing persistence time on DH. As discussed before, the unified scaling description does not seem to be able to accommodate data for large $\tau_p$. So, we separately discuss the effect of large $\tau_p$. In Fig.\ref{fig:largeTaup}(a), we show the block size dependence of $\chi_4^P$ for block size $L_B = L/3$ for various $\tau_p$ ranging from $\tau_p = 0.4$ to $\tau_p = 100$ including passive system. In panel (b), we show the finite size scaling collapse using a correlation length, $\xi(\tau_p)$, which seems to show non-monotonic dependence on $\tau_p$ as shown in panel (c) of the same figure. This non-monotonic dependence of dynamic correlation length with persistence time is indeed very puzzling with no immediate understanding.  These results suggest that further studies are required to understand the dependence of $\chi_4^P$ on $\tau_p$ for various $f_0$, and $c$ systematically to understand whether the growth of correlation length is always non-monotonic for all $f_0$, and $c$. It will indeed be very interesting if a unified scaling description can be derived that could rationalize all the results across a wide variation of parameters $f_0$, $c$, and $\tau_p$ over both glassy and non-glassy regimes.}

\section{Conclusion}
	\SK{To conclude, we have done an extensive analysis of the dynamics of a model active glass-forming liquid over a wide range of temperatures and activity parameters and proposed a scaling theory description to rationalize all the data related to the relaxation process and data for shorter persistence time for the dynamic heterogeneity. In particular, we showed that at large activity limits, the relaxation becomes sub-Arrhenius by crossing over from super-Arrhenius at smaller activities. This crossover process can be rationalized by our proposed scaling theory with a suitable but approximate definition of effective temperature in terms of the three activity parameters, namely active force magnitude $f_0$, concentration of active particles $c$, and persistence time of activity $\tau_p$. By proposing a series representation of effective temperature in terms of scaled effect parameter $\Omega/T$ with $\Omega = cf_0^2\tau_p/(1 + G \tau_p)$ including sub-linear leading order terms, we are able to rationalize the entire spectrum of relaxation process characterized by $\tau_\alpha$ over the entire temperature and activity range. This suggests that effective temperature description can still be a good description of the physics even at large activity where the usual definition of effective temperature \cite{Nandi2018} breaks down. Although this scaling theory describes the data quite well, it does not immediately give us a possible microscopic mechanism for the validity of such a description.     
} 

\SK{Our observation of interesting and completely counter-intuitive finite-size effects in $\tau_\alpha$ at large activity is very intriguing and asks for a possible microscopic origin of such a dependence. This result, we believe, is probably the first observation of its kind in a model glass-forming liquid in the presence of activity. This also indicates a possible crossover in the relaxation mechanisms in these systems from an activated relaxation process at smaller activities to a Mode Coupling-like relaxation process at larger activities. Further studies are needed to better understand this crossover in the finite-size effects in these systems, which may have important implications for making progress in the understanding of glass transition in passive systems.}

\SK{We also demonstrated that the same scaling theory can rationalize the temperature and activity dependence of four-point dynamic susceptibility peak, $\chi_4^P$, in a unified manner for smaller $\tau_p$ values. Note that the dynamic heterogeneity in active systems is very different from its equilibrium counter-part as shown in \cite{Paul2023}; in particular, it was shown that for a given relaxation time or equivalently at a given effective temperature, $\chi_4^P$, the peak height of $\chi_4(t)$, increases with increasing activity strongly when one changes the active force magnitude, $f_0$ in grand-canonical ensemble. In this work, we show that at large activity, $\chi_4^P$ tends to decrease with increasing activity if computed within the canonical ensemble, giving rise to a broad peak-like behavior. Our proposed scaling theory can rationalize all these different behaviors of $\chi_4^P$ with changing activity and temperature in a unified manner. On the other hand, $\chi_4^P$ computed within an effective grand canonical ensemble where all the important components of the fluctuations are included shows a strong growth with increasing activity even at large activity limits in agreement with the observation in \cite{Paul2023} at smaller activities. Finally, we show that the dynamical heterogeneity length shows a non-monotonic growth with increasing $f_0$, eventually saturating to a constant at large enough activity in complete agreement with the finite-size effects observed in $\chi_4^P$. Dynamic heterogeneity at large persistence time shows very interesting behavior that can not be understood using the same scaling theory, but if one analyses these data, one finds a non-monotonic growth of correlation length with increasing persistence time. These results are very counter-intuitive and ask for further detailed studies for better understanding.    
}

\SK{Finally, we believe that our scaling theory description of the relaxation process in active glass-forming liquid over a wide range of temperatures and activity will lead to further refinement of the existing theories of active glasses and hopefully spur more future studies to better understand the fascinating dynamical behavior of active glasses.}

\begin{acknowledgments}
\SK{SK would like to acknowledge funding by intramural funds at TIFR Hyderabad from the Department of Atomic Energy (DAE) under Project Identification No. RTI 4007. SK acknowledges the generous support from the Science and Engineering Research Board (SERB) via Swarna Jayanti Fellowship grants DST/SJF/PSA-01/2018-19 and SB/SFJ/2019-20/05. SK also acknowledges research support from MATRICES Grant MTR/2023/000079 from SERB.}
\end{acknowledgments}



\bibliographystyle{apsrev4-2}
\bibliography{highActivityGlass}

\end{document}